\documentclass[aps, prc, reprint, superscriptaddress]{revtex4-1}
\usepackage{amsmath}
\usepackage{amsfonts}
\usepackage{amssymb}
\usepackage{dcolumn}
\usepackage{lineno}
\usepackage{multirow}
\usepackage{array}
\usepackage{graphicx}
\usepackage[caption=false]{subfig}
\usepackage[colorlinks]{hyperref}
\hypersetup{linkcolor=blue, citecolor=blue, filecolor=blue, urlcolor=blue}
\usepackage[all]{hypcap}
\usepackage{color}
\usepackage[utf8]{inputenc}
\hypersetup{%
    ,urlcolor=blue
    ,citecolor=blue
    ,linkcolor=blue}

\begin{document}

\title{UrQMD Simulations of Higher-order Cumulants in Au+Au Collisions at High Baryon Density}

\author{Xin Zhang}
\affiliation{Institute of Modern Physics, Chinese Academy of Sciences, Lanzhou 730000, China}

\author{Yu Zhang}
\affiliation{School of Physics and Technology, Guangxi Normal University, Guilin 541001, China}

\author{Xiaofeng Luo}
\affiliation{Key Laboratory of Quark and Lepton Physics (MOE) and Institute of Particle Physics,Central China Normal University, Wuhan 430079, China}

\author{Nu Xu}
\affiliation{Institute of Modern Physics, Chinese Academy of Sciences, Lanzhou 730000, China}
\affiliation{Key Laboratory of Quark and Lepton Physics (MOE) and Institute of Particle Physics,Central China Normal University, Wuhan 430079, China}


\date{\today}


\begin{abstract}
High moments of conserved quantities such as net-baryon, net-electric charge, and net-strangeness in heavy-ion collisions are sensitive to fluctuations caused by the QCD critical point (CP). The event-by-event analysis of high moments of the conserved charges has been widely used in experiments to search for the CP, especially in the RHIC-STAR experiment.

In order to establish a {\it dynamical non-critical base line}, especially at the high baryon density region, we have performed a systematic analysis of the proton multiplicity distributions from Au+Au collisions at 3 $\leq$ $\sqrt{s_{NN}}$ $\leq$ 9.2 GeV collisions. The results on beam energy,  centrality and rapidity width dependence of proton (factorial) cumulants, up to the $4^{th}$ order,  are extracted from the calculations of the hadronic transport model UrQMD. In addition, the effects of initial volume fluctuation is also discussed. These results will be important when we do physics analysis the RHIC beam energy scan (BES) data, especially for the fixed-target data and experimental data from future CBM experiment at FAIR.
\end{abstract}

\maketitle


\section{Introduction}
The main task of heavy-ion collision experiments is to study the properties of Quark-Gluon Plasma (QGP) and the phase structure of strongly interacting matter, which can be displayed by a QCD (Quantum Chromodynamics) phase diagram in terms of temperature $T$ and baryon chemical potential ($\mu_{B}$). Recently, theoretical studies have predicted a QCD critical point at the end of the first-order phase transition boundary while there is no consensus on the theoretical side on the exact location of the CP~\cite{Clarke:2024ugt,Basar:2023nkp,Hippert:2023bel,Fu:2019hdw,Gunkel:2021oya,Gao:2020fbl,Sorensen:2024mry,Shah:2024img}.

On the experimental side, in the search for the QCD critical point, higher-order cumulants of conserved quantities such as net-baryon, net-strangeness, and net-electric charge are proposed as promising observables due to their sensitivity to system correlation length $\xi$, which tends to diverge at around critical point~\cite{Hatta:2003wn,Asakawa:2015ybt,Kitazawa:2017ljq,Luo:2017faz}. The Beam Energy Scan program at the BNL Relativistic Heavy Ion Collider (RHIC) by the STAR experiment has collected large datasets over a wide range of collision energies $\sqrt{s_{NN}}$ = 7.7 -- 200 GeV of Au+Au collisions. Assuming that the thermalization is reached in those collisions, the corresponding range of chemical potential is $\mu_B \sim$ 420 -- 25 MeV. The STAR Experiment has published a series of papers on high moments related to the QCD critical point~\cite{STAR:2010mib, STAR:2013gus, STAR:2014egu, STAR:2017tfy,
STAR:2019ans,STAR:2020tga,STAR:2021iop,STAR:2021rls,STAR:2021fge,STAR:2022etb,STAR:2022vlo,STAR:2023zhl,STAR:2023ebz}. Recent STAR results on the net-proton $C_{4}/C_{2}$~\cite{STAR:2020tga,STAR:2021iop,Chen:2024aom} show a non-monotonic energy dependence trend with a significance level of 3.1$\sigma$, which could be a signature of QCD CP. The subsequent net-proton analysis on BES-II datasets ($\sqrt{s_{NN}}$ = 7.7 -- 27 GeV), which has 10 to 20 times larger statistics
has significantly reduced statistical uncertainties~\cite{STAR:2025zdq}. During the second phase of the beam energy scan, BES-II, STAR also took data in fixed-target mode (FXT) and has collected Au+Au collisions data at $\sqrt{s_{NN}}$ = 3.0 -- 4.5 GeV for proton high-moments analysis. The FXT data has extended the QCD CP search to unprecedented high baryon density region where $\mu_B \sim$ 750 MeV. The first publication of proton high moments of STAR fixed-target data in Au+Au collisions at $\sqrt{s_{NN}}$ = 3.0 GeV shows the consistency between data and hadronic transport model calculation indicating that the critical point could exist only at energies higher than 3.0 GeV~\cite{STAR:2021fge, STAR:2022etb}. Measurements at baryon density region will be crucial to establish a complete energy dependence trend. There have been many publications of non-critical model calculations~\cite{Xu:2016qjd} as baselines for net-proton high moment measurement while few in high baryon density region. As noted in Refs.~\cite{STAR:2022etb}, fluctuation analysis suffers from the so-called initial volume fluctuation effect. The effect emerges in experimental measurements when using charged particle multiplicity as a reference of collision centrality. From the experimental side, one could estimate the initial volume fluctuation effect by evaluating the centrality resolution of reference multiplicity at a certain energy. Generally, limited reference multiplicity means low centrality resolution which brings a large initial volume effect into high moment measurement, especially for higher-order ones.

In this paper, we show (net-)proton moments calculations using the hadronic transport model UrQMD (Ultra-relativistic Quantum Molecular Dynamics) for Au+Au collisions at $\sqrt{s_{NN}}$ = 3.0 -- 9.2 GeV, and discuss the effect of initial volume fluctuation on high moments. In Sec.~\ref{sec:setup}, we briefly introduce the UrQMD model and simulation setup, as well as calculation details. In Sec.~\ref{sec:result}, we show the model results and compare with experimental data. We summarize in Sec.~\ref{sec:summary}.

\section{The UrQMD model and calculation details}\label{sec:setup}
\subsubsection{The UrQMD Model}

The UrQMD model~\cite{Bass:1998ca,Bleicher:1999xi} is a hadronic transport model that is successfully used in simulating (ultra-)relativistic heavy-ion collisions at a wide range of energies from SIS to RHIC. In UrQMD, hadrons have explicit space-time evolution trajectories. Below $\sqrt{s_{NN}}$ = 5 GeV, particle production is described by interactions between hadrons and resonances, and at collision energies above $\sqrt{s_{NN}}$ = 5 GeV, the excitation of color strings and their fragmentation into hadrons dominates particle production.

In this paper, we simulate Au+Au collisions at $\sqrt{s_{NN}}$ = 3.0 -- 9.2 GeV in UrQMD configured as the standard cascade mode. Impact parameter (b) is set to 0 to 16 fm. The statistics at each energy are shown in Table~\ref{tab_stat}.

\begin{table}[!ht]\label{tab_stat}
\center
\caption{Statistics of Au+Au collisions simulated in UrQMD model at $\sqrt{s_{NN}}$ = 3.0 -- 9.2 GeV}
\begin{tabular}{c|c|c|c|c|c|c|c|c}
\toprule
    $\sqrt{s_{NN}}$ (GeV) & 3.0 & 3.2 & 3.5 & 3.9 & 4.5 & 4.9 & 7.7 & 9.2  \\  \hline
    Events (million)      & 66  & 117 & 131 & 105 & 93  & 136 & 170 & 156  \\ \hline
\end{tabular}
\end{table}

\subsubsection{Analysis Details}
Cumulants of a distribution up to fourth order are defined as follows: 
\begin{equation}
\begin{split}
    C_{1} &= \langle N \rangle, \\
    C_{2} &= \langle(\delta N)^2\rangle, \\
    C_{3} &= \langle(\delta N)^3\rangle, \\
    C_{4} &= \langle(\delta N)^4\rangle - 3\langle(\delta N)^2\rangle^2,
\end{split}
\end{equation}
where N is the number of proton in each event and $\delta N = N - \langle N\rangle$ is the deviation from mean value $\langle N\rangle$. Cumulant ratios such as $C_{2}/C_{1}$, $C_{3}/C_{2}$, and $C_{4}/C_{2}$ are used to cancel volume dependence and are shown below:

\begin{equation}
\begin{split}
\frac{C_{2}}{C_{1}} = \sigma^2/M, \enspace\frac{C_{3}}{C_{2}} = S\sigma, \enspace\frac{C_{4}}{C_{2}} = \kappa\sigma^2
\end{split}
\end{equation}

Where M, $\sigma^2$, S, $\kappa$ are known as mean value, variance, skewness, and kurtosis, respectively. Higher-order cumulants such as skewness and kurtosis are famous for their capability to measure non-Gaussianity. The factorial cumulants ($\kappa_{n}$) are related to the corresponding cumulants through the following relations:
\begin{equation}
\begin{split}
    \kappa_{1} &= C_{1} = \langle N \rangle, \\
    \kappa_{2} &= -C_{1} + C_{2}, \\
    \kappa_{3} &= 2C_{1} - 3C_{2} + C_{3}, \\
    \kappa_{4} &= -6C_{1} + 11C_{2} - 6C_{3} + C_{4}
\end{split}
\end{equation}

In this paper, the calculations of proton and net-proton high moments follow the same procedure used in experimental data analysis. Collision centrality is determined using charged particle multiplicity including $\pi^{\pm}$ and $K^{\pm}$. To approximate the acceptance of the STAR Au+Au collisions experiment at each energy (3 -- 9.2 GeV), $\pi^{\pm}$ and $K^{\pm}$ are counted within $-2.0 < \eta < 0$ for 3.0 GeV, $-2.4 < \eta < 0$ for 3.2 -- 4.9 GeV, and $|\eta| < 1.6$ for 7.7 and 9.2 GeV. Protons and anti-protons are not included in the multiplicity definition to avoid self-correlation~\cite{Luo:2013bmi,Chatterjee:2019fey}. 
The formation of light fragments such as deuterons is not included in the UrQMD calculation. As shown in previous studies~\cite{PhysRevC.98.054620,Ye:2020lrc}, Deuterons and other light nuclei are formed via the coalescence of nearby protons and neutrons, which effectively removes correlated protons from the free nucleon sample and leads to a increase of the proton cumulant ratios. As discussed in~\cite{Chatterjee:2020nnn}, they showed the effect of deuteron formation is similar to the binomial efficiency effect due to the loss of protons via deuteron formation. The formation of light fragments should be taken into account when comparing with data.

\begin{figure}[htbp]
    \centering
    \includegraphics[width=0.95\linewidth]{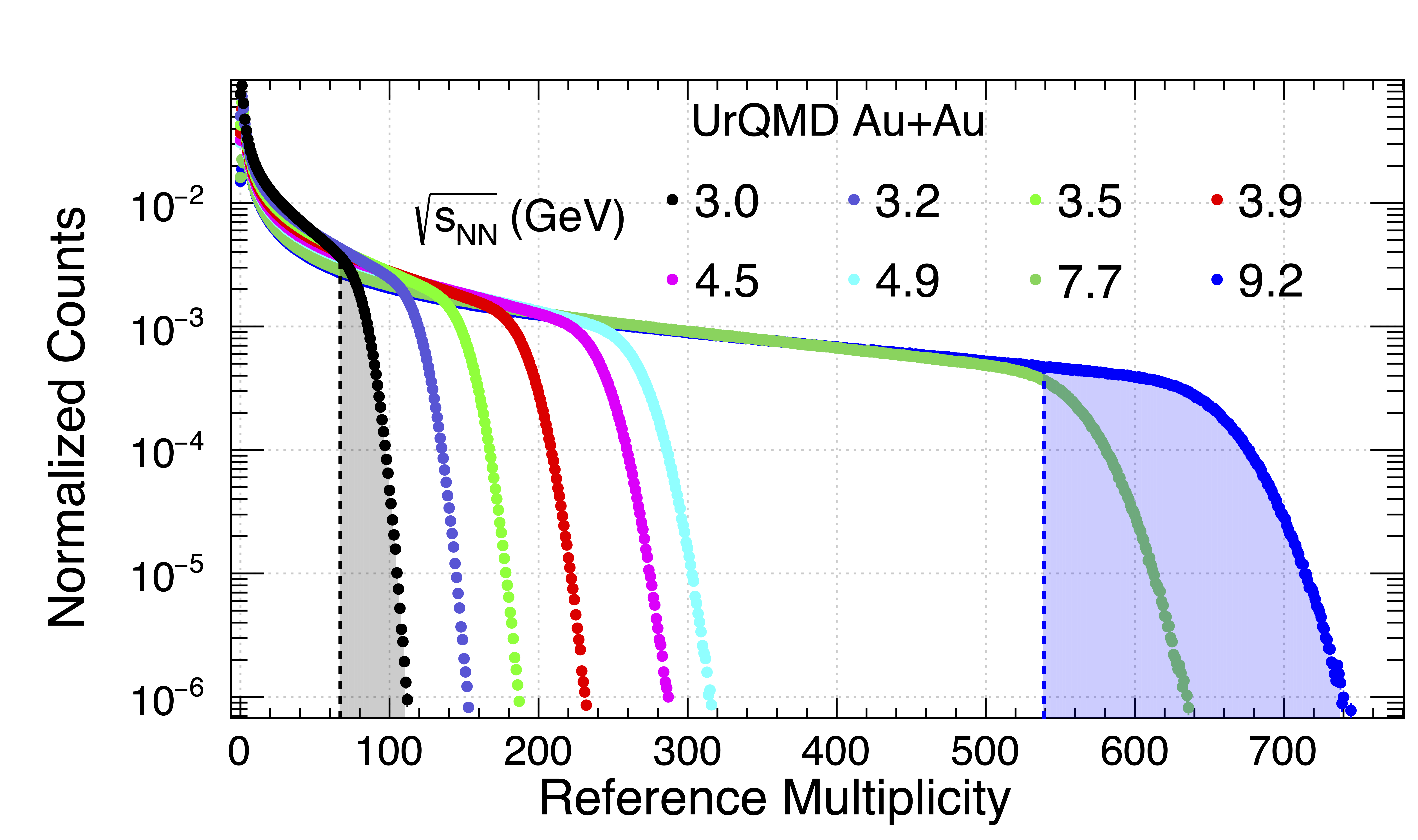}
    \caption{Reference multiplicity distribution in Au+Au collisions at $\sqrt{s_{NN}}$ = 3.0, 3.2, 3.5, 3.9, 4.5, 4.9, 7.7 and 9.2 GeV  calculated by the UrQMD model. The regions with black and blue shading indicate the most central 0--5\% collision.}
    \label{fig_ref3}
\end{figure}

Figure~\ref{fig_ref3} shows charged particle multiplicity distributions in Au+Au collisions at $\sqrt{s_{NN}}$ = 3.0 -- 9.2 GeV in UrQMD model. The charged particle multiplicity distribution is then fitted by the Monte Carlo Glauber model~\cite{Miller:2007ri}, and categorized into several centrality classes: 0--5\%, 5--10\%, $\cdots$, 70--80\%. The black and blue shaded regions of Fig.~\ref{fig_ref3} indicate the 0--5\% collision centrality classes. Centrality bin width correction (CBWC) method~\cite{Luo:2013bmi} is used in proton and net-proton cumulant calculation. Cumulants are calculated in each reference multiplicity bin and then taken weighted average for each centrality bin. The statistical errors are determined by analytical equations using the Delta method~\cite{Luo2014}. 

\begin{figure}[htbp]
    \centering
    \includegraphics[width=0.9\linewidth]{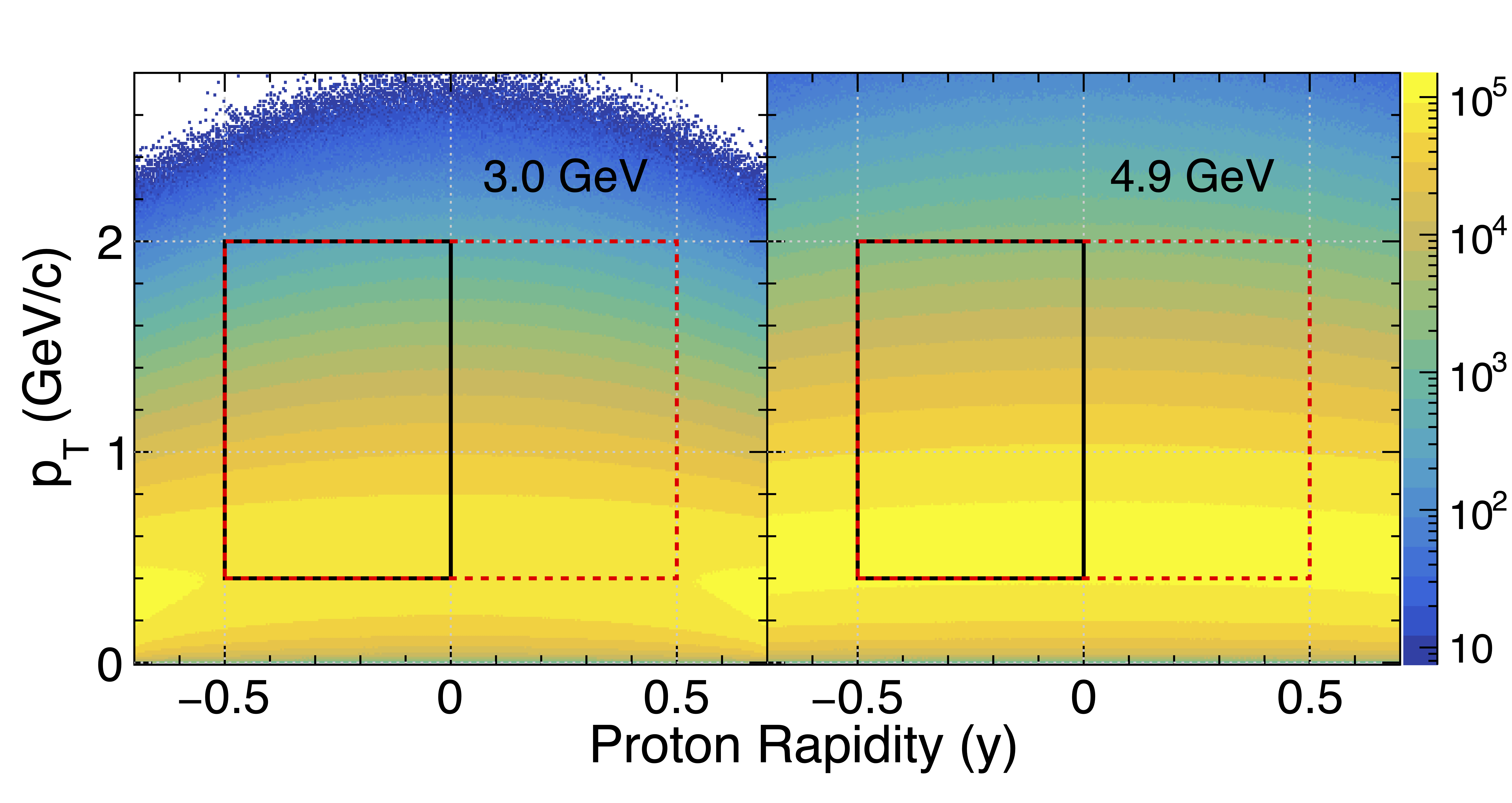}
    \caption{Transverse momentum ($p_{\rm T }$) versus proton rapidity (y) in Au+Au minimum bias collisions at $\sqrt{s_{NN}}$ = 3.0 and 4.9 GeV calculated by the UrQMD model. Protons from rapidity window $|y| < 0.5$ (red-dashed box) and $-0.5 < y < 0.0$ (black box) are used in the analysis for the collider and FXT mode, respectively.}
    \label{fig_phase}
\end{figure}

Figure~\ref{fig_phase} shows the proton acceptance in transverse momentum ($p_{\rm T }$) versus proton rapidity ($y$) in center-of-mass frame in Au+Au collisions at $\sqrt{s_{NN}}$ = 3.0 and 4.9 GeV from UrQMD model. The black and red boxes indicate analysis windows within $0.4 < p_{T} < 2.0$ GeV/$c$ and $-0.5 < y < 0$ and $|y|<0.5$.


\section{Results and Discussion}\label{sec:result}

\begin{figure}[htbp]
    \centering
    \includegraphics[width=0.9\linewidth]{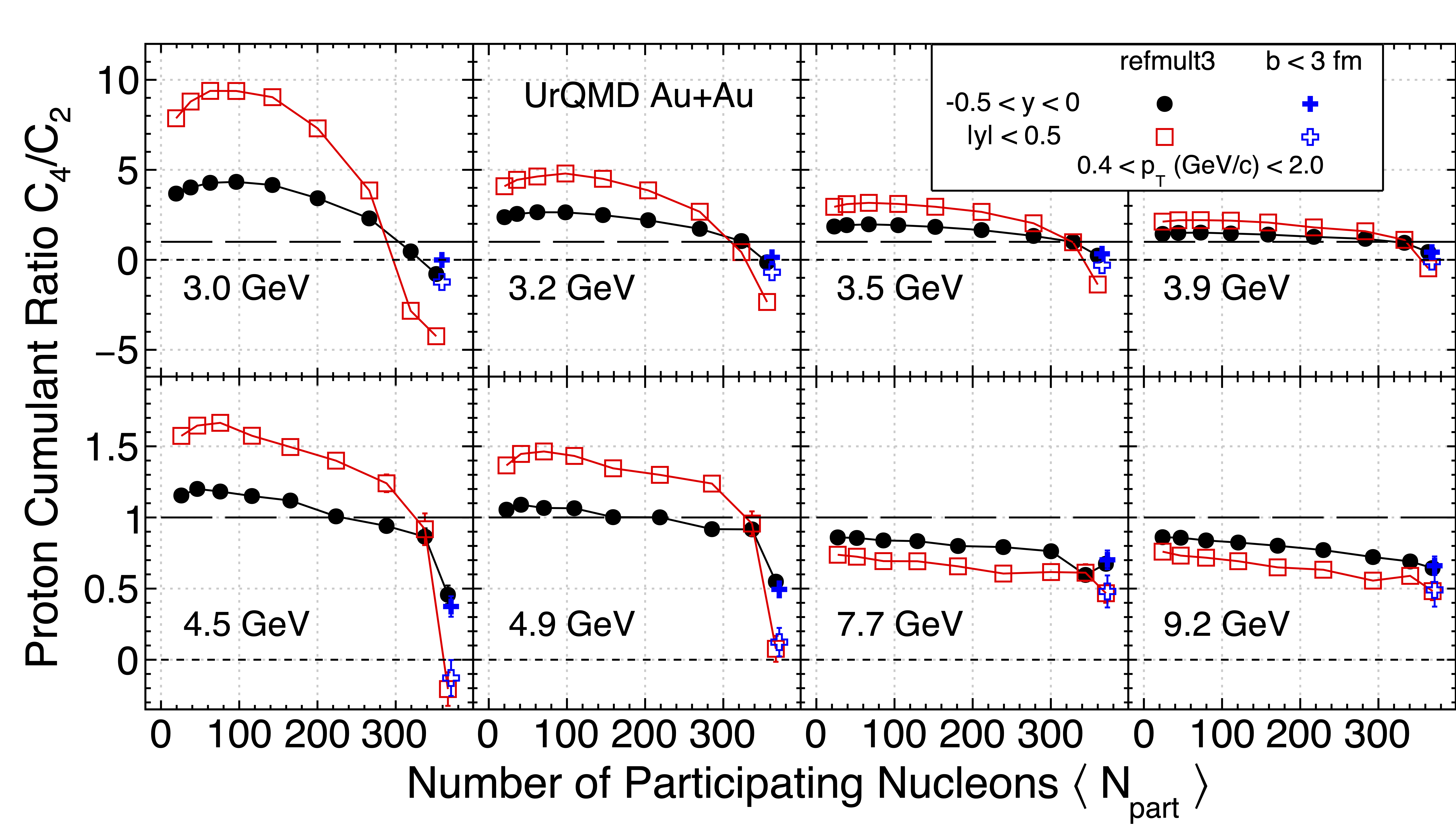}
    \caption{Centrality dependence of the proton cumulant ratio, $C_{4}/C_{2}$, in Au+Au collisions at $\sqrt{s_{NN}}$ = 3.0 -- 9.2 GeV calculated by the UrQMD model. Solid black circles and red open squares represent results for proton in the rapidity acceptance -0.5 $<$ y $<$ 0 and $\left|y\right|$ $<$ 0.5, respectively. Blue solid crosses and open crosses are calculations with a cut on impact parameter $b\leq$ 3 fm, for -0.5 $<$ y $<$ 0 and $\left|y\right|$ $<$ 0.5, respectively. For these calculations, the $p_{\rm T}$ acceptance of protons and anti-protons is $0.4 < p_{T} < 2.0$ GeV/$c$.}
    \label{fig_cent}
\end{figure}

\begin{figure}[htbp]
    \centering
    \includegraphics[width=0.9\linewidth]{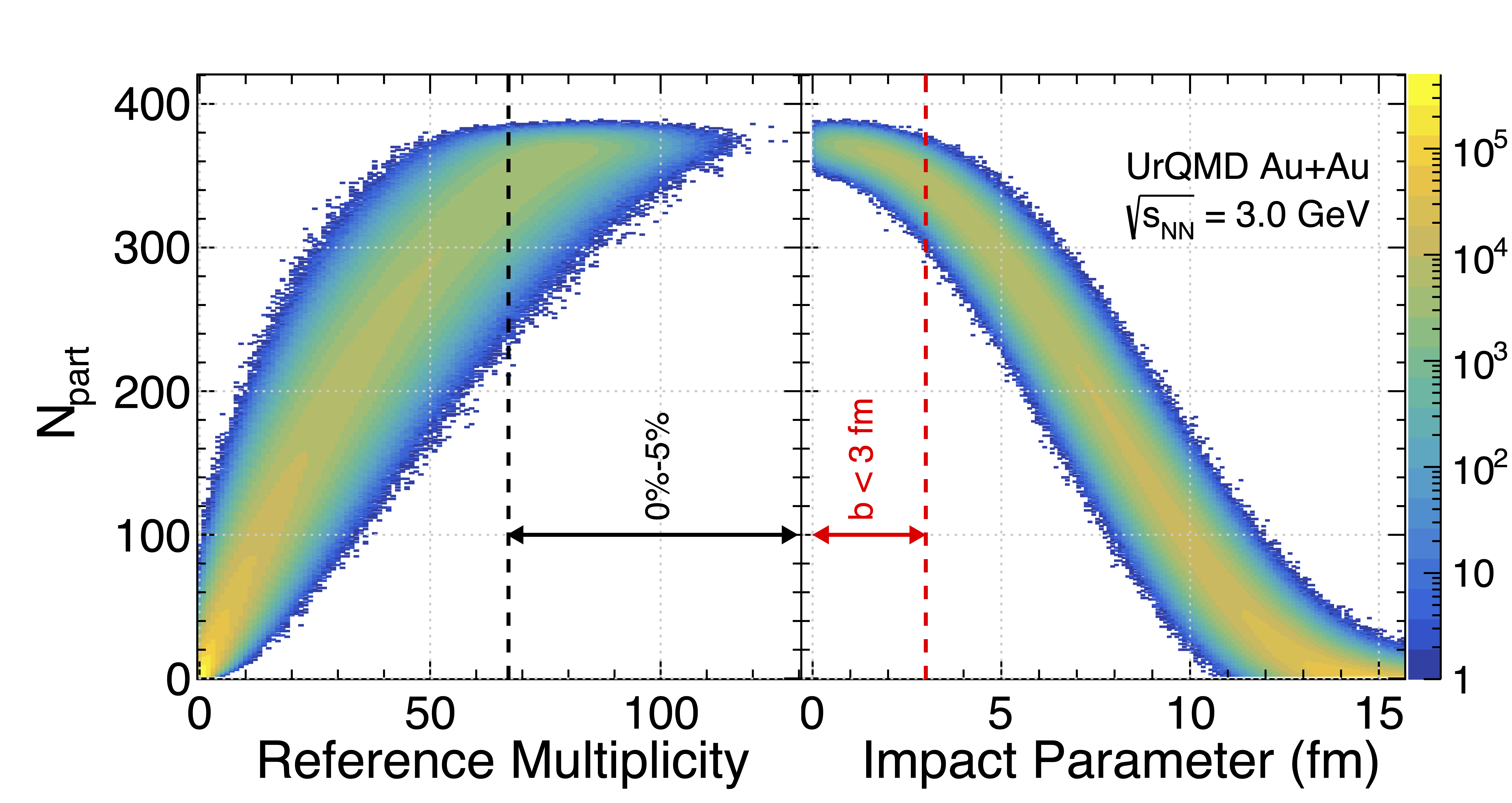}
    \caption{Left panel: Correlation distribution of reference multiplicity vs. $N_{part}$ in Au+Au collisions at $\sqrt{s_{NN}}$ = 3.0 GeV calculated by the UrQMD model. The vertical black dashed line indicate the 0--5\% central collisions selected by reference multiplicity. Right panel: Correlation distribution of b vs. $N_{part}$. The vertical red dashed line indicate a cut on b $\leq$ 3 fm.}
    \label{fig_corr}
\end{figure}

In this section, we show the calculations of proton and net-proton cumulant ratios in UrQMD model.

Figure~\ref{fig_cent} shows centrality dependence of proton $C_{4}/C_{2}$ in Au+Au collisions at $\sqrt{s_{NN}}$ = 3.0 -- 9.2 GeV calculated by the UrQMD model.
The dashed line at unity indicates a baseline from Poisson distribution. In mid-central and peripheral collisions, the proton $C_{4}/C_{2}$ largely deviate from Poisson baseline, which demonstrates poor centrality resolution of reference multiplicity especially in lower energies.
The proton $C_{4}/C_{2}$ shown with black dots are closer to Poisson baseline than results shown with red squares, which could be explained by that the narrower proton distribution with $-0.5<y<0$ converges to Poisson statistic rather than a wider distribution selected by $|y|<0.5$. 


As one can see in the figure, the proton high moments, $\kappa\sigma^{2}$, have been suppressed in both central and peripheral collisions, especially at the lowest collision energy $\sqrt{s_{NN}}$ = 3.0 GeV. The observed suppression decreases with increasing collision energy, and is more pronounced for the wider rapidity $|y|<0.5$. In addition, the largest fluctuation from unity is seen in the mid-central collisions $\langle N_{part} \rangle \sim 100$. The negative value of the ratios is caused by the initial volume fluctuation, see discussions below.  The suppression at the most central collisions is due to the fixed number of incoming nucleons. 

Now, let us focus on the results from the most central collisions: for protons from rapidity $|y|<0.5$, the values of $C_{4}/C_{2}$ ratios are negative till the collision energy reaches 4.9 GeV and this is true even for events selected from the cuts by the impact parameter $b$ as shown by the open- and filled-blue crosses in the figure. For collision energy  $\sqrt{s_{NN}} >$ 4.5 GeV, all ratios become positive, the values of the ratios from wider rapidity bins become smaller than those from narrower bin $-0.5<y<0$ and they are decreased from peripheral to central collisions. All of these indicated the diminishing role of the initial volume fluctuations and baryon number conservations in such collisions.

Figure~\ref{fig_corr} shows the correlation distribution of reference multiplicity vs. $N_{part}$ and b vs. $N_{part}$ in Au+Au collisions at $\sqrt{s_{NN}}$ = 3.0 GeV, calculated by the UrQMD model. The vertical black dashed line indicate the 0--5\% central collisions selected by reference multiplicity, and the vertical red dashed line indicate a cut on $ b \leq$ 3 fm. One can compare these two cuts on the $N_{part}$ distributions. As one can see, the $\langle N_{part} \rangle$ determined by reference multiplicity, as it is done in experiment, shows a much wider variation in the $N_{part}$ distribution, see left plot,  compared to that of the $\langle N_{part} \rangle$ extracted from the cut on the impact parameter $b \leq$ 3 fm, see right plot. Reversely, for a fixed range of the $N_{part}$ distribution, a much wider fluctuation is seen in the 'measured' reference multiplicity compared to the impact parameter. The variations in the $N_{part}$ distribution are the root cause of the initial volume fluctuations shown in Figure~\ref{fig_cent}.
One should note that such variation is part of the collision dynamics and it should be properly simulated in the physics analysis.

\begin{figure}[htbp]
    \centering
    \includegraphics[width=0.9\linewidth]{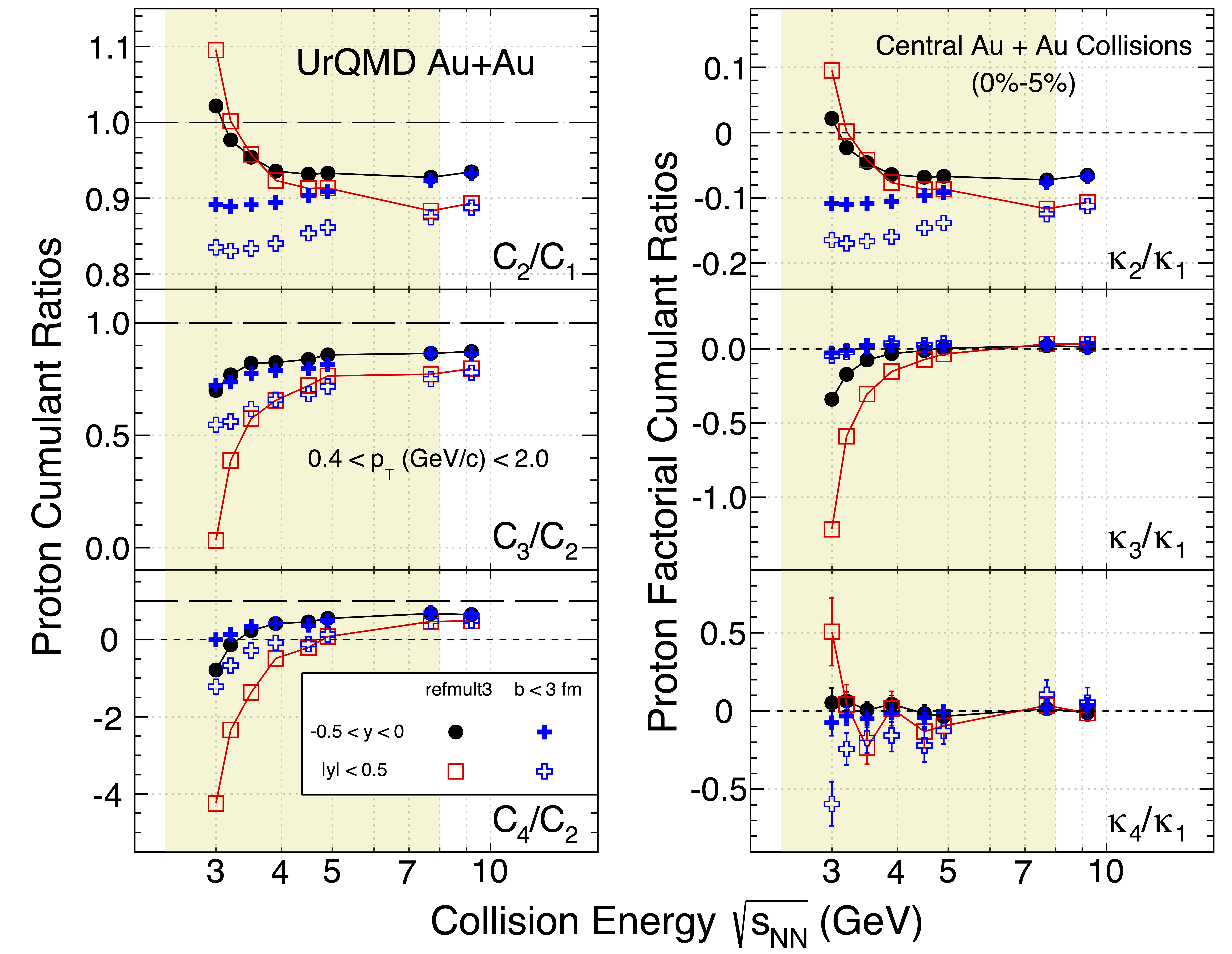}
    \caption{Collision energy dependence of proton cumulant ratios ($C_{2}/C_{1}$, $C_{3}/C_{2}$ and $C_{4}/C_{2}$) and factorial cumulant ratios ($\kappa_{2}/\kappa_{1}$, $\kappa_{3}/\kappa_{1}$ and $\kappa_{4}/\kappa_{1}$) in 0--5\% central Au+Au collisions from the UrQMD model. Black solid circles and red open squares represent proton results in the rapidity acceptance -0.5 $<$ y $<$ 0 and $\left|y\right|$ $<$ 0.5, respectively. Blue solid crosses and open crosses are calculations with a cut on b $\leq$ 3 fm, for -0.5 $<$ y $<$ 0 and $\left|y\right|$ $<$ 0.5, respectively.}
    \label{fig_ratios}
\end{figure}

\begin{figure}[htbp]
    \centering
    \includegraphics[width=0.9\linewidth]{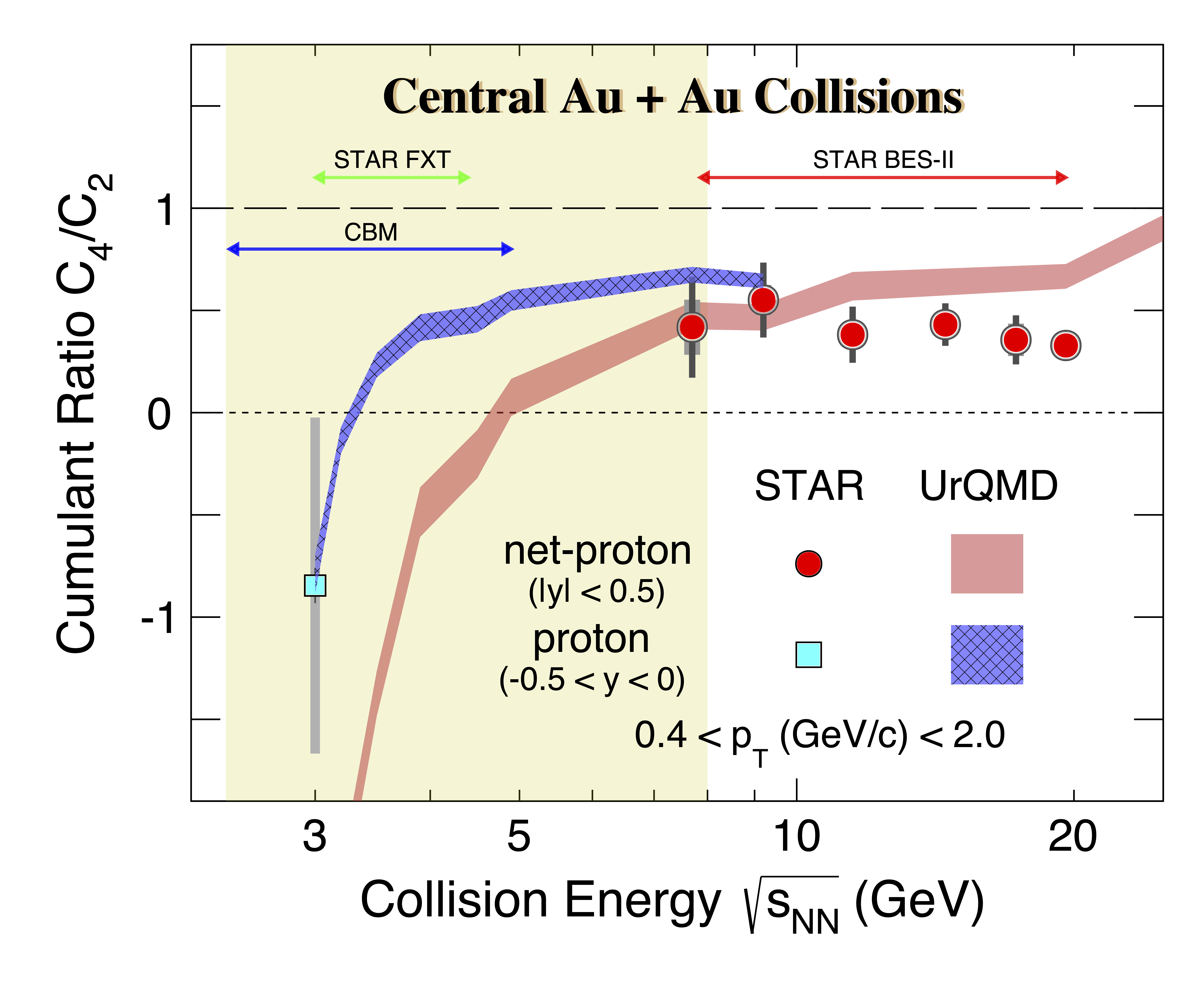}
    \caption{Collision energy dependence of proton and net-proton cumulant ratio, $C_{4}/C_{2}$, in 0--5\% central Au+Au collisions from the UrQMD model. The blue and red band represent proton and net-proton results in UrQMD, respectively.
    The red and cyan markers represent STAR measurements, red from BES-II, cyan form fixed-target experiment.
    In addition, the green and blue arrows indicate collision energy range of STAR fixed-target experiment at RHIC and CBM experiment at FAIR, respectively.}
 \label{fig_r42}
\end{figure}

Figure~\ref{fig_ratios} depicts proton cumulant ratios $C_{2}/C_{1}$, $C_{3}/C_{2}$ and $C_{4}/C_{2}$ (left panels) and factorial cumulant ratios $\kappa_{2}/\kappa_{1}$, $\kappa_{3}/\kappa_{1}$ and $\kappa_{4}/\kappa_{1}$ (right panels) as a function of collision energy in central collisions 0--5\% within kinematic acceptance 0.4 $<$ $p_{\rm T}$ $<$ 2.0 GeV/$c$. Results of b $\leq$ 3 fm are shown as solid cross (-0.5 $<$ y $<$ 0) and open cross ($\left|y\right|$ $<$ 0.5). As mentioned in the previous paragraph, the large difference between the reference multiplicity 0--5\% and b $\leq$ 3 fm is mainly caused by initial volume fluctuations. As the collision energy increases, While all $C_{3}/C_{2}$, $C_{4}/C_{2}$ and $\kappa_{3}/\kappa_{1}$ ratios show a initial fast increase and saturation at energy above 5 GeV, the ratios of $C_{2}/C_{1}$ and $\kappa_{2}/\kappa_{1}$ just behave in the opposite way. In any case, protons 
from wider rapidity bins show more sensitivity to initial volume fluctuation especially at the high baryon density region.


Figure~\ref{fig_r42} depicts the collision energy dependence of the proton and net-proton $C_{4}/C_{2}$ ratios from 0--5\% central collisions. The experimental measurements in $\sqrt{s_{NN}}$ = 7.7 -- 19.6 GeV Au+Au collisions from RHIC-STAR BES-II data are also shown as red solid circles~\cite{STAR:2020tga} from collider collisions and the 3 GeV result is shown as blue square. Results of UrQMD model calculations are displaced as colored bands: protons from $-0.5 < y < 0$ and net-protons from $|y|<0.5$ are presented as blue and red band, respectively. In both cases, both bands decreased as collision energy decreases due to both baryon number conservation and volume fluctuations. Due to large acceptance in case of the net-protons from $|y|<0.5$, the decreases of the $C_{4}/C_{2}$ ratios are much faster at the low energy region. While large deviation between data and UrQMD calculation at the 19.6 GeV, the transport model well reproduced the data at $\sqrt{s_{NN}}$ = 3 GeV. The result implies that if the critical point exists, it should be somewhere between $3 < \sqrt{s_{NN}} <$ 19.6 GeV Au+Au collisions. In addition to the beam energy scan program at RHIC, the physics program at the FAIR CBM experiment~\cite{CBM_FAIR} is essential in order to finally determine the location of the QCD critical point. 

\section{Summary}\label{sec:summary}
In summary, we presented results of proton (from $-0.5 < y < 0$) and net-proton (from $|y|<0.5$) high order cumulants ratios from Au+Au collisions at $\sqrt{s_{NN}}$ = 3.0 -- 19.6 GeV from the hadronic transport model UrQMD calculations. These results are valuable dynamic references for the QCD critical point search and there are two main observations from these analyses: (i) At low energy  $\sqrt{s_{NN}} <$ 5 GeV, initial volume fluctuation is important, the larger the acceptance the stronger the effect of the fluctuation and (ii) In the FXT Au+Au collisions below $\sqrt{s_{NN}} <$ 5 GeV, UrQMD model calculations well reproduced $C_{4}/C_{2}$ ratios implying that if the QCD critical point exist, it should be in the energy $\sqrt{s_{NN}} >$ 3.0 GeV.  

Recently, preliminary results of proton cumulant on RHIC-STAR fixed-target experiment at $\sqrt{s_{NN}}$ = 3.2 -- 3.9 GeV are reported~\cite{Sweger:2025qm}. The future Compressed Baryonic Matter (CBM) experiment in Facility for Antiproton and Ion Research (FAIR)~\cite{CBM_FAIR} will cover a collision energy of $\sqrt{s_{NN}}$ = 2.4 -- 4.9 GeV, with excellent acceptance and higher statistics to reduce both statistical and systematic uncertainties, will play an important role in the QCD critical point search. 



\bibliography{reference}

\end{document}